\documentclass[aps,prl,twocolumn,superscriptaddress]{revtex4-1} 

\usepackage{graphicx}
\graphicspath{{C:/Users/LimbeekMAJ/Dropbox/EMS/paperbubblecoolingH20/figures/}}
\usepackage{hyperref}
\usepackage{siunitx}
\usepackage{setspace}

\begin{document}
\title{Mechanics of cooling liquids by forced evaporation in bubbles}

\author{Michiel A. J. van Limbeek}
\affiliation{Energy Materials and Systems Group, Department of Science and Technology, University of Twente, 7500 AE Enschede, The Netherlands}
\affiliation{Physics of Fluids Group, Max Planck Center for Complex Fluid Dynamics, J. M. Burgers Center for Fluid Dynamics and MESA+ Research Institute, Department of Science and Technology, University of Twente, P.O. Box 217, 7500 AE Enschede, The Netherlands}

\author{D. van Buuren}
\affiliation{Energy Materials and Systems Group, Department of Science and Technology, University of Twente, 7500 AE Enschede, The Netherlands}
\author{M.R.P. van den Broek}
\affiliation{Energy Materials and Systems Group, Department of Science and Technology, University of Twente, 7500 AE Enschede, The Netherlands}
\author{H.J.M. ter Brake}
\affiliation{Energy Materials and Systems Group, Department of Science and Technology, University of Twente, 7500 AE Enschede, The Netherlands}
\author{S. Vanapalli}
\affiliation{Energy Materials and Systems Group, Department of Science and Technology, University of Twente, 7500 AE Enschede, The Netherlands}


\begin{abstract}
 Injecting a non-dissolvable gas into a saturated liquid results in sub-cooling of the liquid due to forced evaporation into the bubble. Previous studies assumed the rate of evaporation of liquid into the bubble to be independent of the degree of sub-cooling. In our study we quantify  the bubble growth by direct observation using high speed imaging and prove that this hypothesis is not true. A phenomenological model of the bubble growth as a function of the degree of sub-cooling is developed and we find excellent agreement between the measurements and theory. This bubble cooling process is employed in cooling a liquid. By identification of all heat flows, we can well describe the cool down curve using bubble cooling. Bubble cooling provides an alternative cooling method for liquids without the use of complicated cooling techniques.
\end{abstract}
\maketitle
Evaporation of liquids is an efficient way to transfer heat away from a hot solid, a method employed frequently in the context of heat-pipes, power-plants and industrial spray cooling. A daily-life example is the use of transpiration to regulate our body temperature. The amount of liquid which can evaporate into vapour depends largely on the liquid's temperature and its vapour pressure. When injecting a bubble of a non-condensible substance  into a liquid bath, this partial vapour pressure is initially zero, resulting in an immediate evaporation and hence, cooling of the liquid.

In some cases this effect is very desirable. For example, cryogenic liquids are often at their saturation temperature following their liquefaction. It was found that bubbling helium gas into a liquid nitrogen tank lowers the temperature  of the liquid by more than ten Kelvin and suppresses (vapour) bubble generation \cite{Minkoff1957,Schmidt1963, Lytle1965, Xu1993,Takayoshi2009}. Firstly, subcooling the liquid allows one to perform studies in cryogenic liquids without any disturbances caused by vapour bubbles \cite{townsend2016}. Another advantage is that the cooling rate of quencing procedures is greatly enhanced when the liquid is subcooled, which is of great importance in the context of cryo-preservation \cite{leek2006frozen,morente2006}.  Furthermore, in contrast to  depressurizing a tank to achieve a (temporary) lower saturation temperature and then increasing the pressure again, this bubble cooling method does not require the system to withstand reduced pressures and it will offer a continuous rather than a batch sub-cooling of the liquid.

However, this cooling is not favourable for all systems. Often this is a side-effect of another process which leads to efficiency losses. As an example, in a bubble column reactor the injected gas is to react with the liquid \cite{shimizu2000} or to remove already dissolved gasses (sparging) \cite{brown1981}. In both cases, the change from the liquid into the gaseous state will involve an energy penalty, in the form of latent heat, cooling down the liquid phase. Moreover, in the case of sparging, a fraction of the liquid will be removed as well by the bubble process.

\begin{figure}
\centering
\includegraphics[scale=0.8]{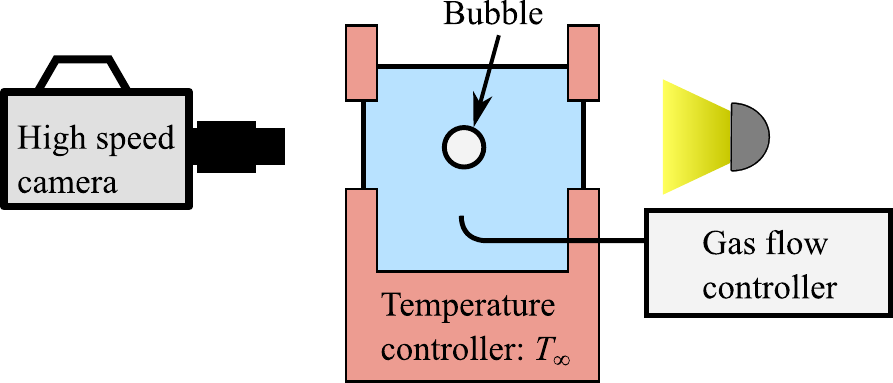}
\caption{Schematic of the setup to measure the bubble growth. Most components described in the text are omitted for clarity.}
\label{fig:SetupBare}
\end{figure}

For both cases, whether the cooling being desirable (bubble cooling) or not (bubble columns), little is known on how much energy and mass is exchanged during this process. 
In this study, we optically address these questions for a cryogenic system, where helium gas is injected into liquid nitrogen, and a non-cryogenic case in which nitrogen gas is fed into a water bath. We focus on how the bubbles grow by forced evaporation, depending on the temperature of the liquid bath. We propose a  theoretical model of this cooling mechanism and verify the model by optical observations. Secondly, we characterize the cooling of a known system, using forced evaporation, where we apply the results of the first part.
The growth of the bubbles was measured using a setup, which had in essence only four major components: a temperature controlled container, a high speed camera, a light source and a bubble injector, see \autoref{fig:SetupBare}. A transparent container was used for optical access, which was equipped with a temperature controller.  We used a Photron APX-RS high speed camera, operating at \SI{1000}{fps} and a shutter speed of \SI{0.5}{\milli \second} to prevent motion blurring. A K2 Distamax lens-system was used to have a field of view of roughly \SI{12}{\milli \metre} wide. A fiber Schott lamp with a diffuser was used as a light source. 
The gas was injected using an upright needle and controlled by a mass-flow controller. The flow rate was in the order of \SI{e-2}{\milli \gram \per \second} to ensure sufficient spacing between the bubbles to be able to perform accurate bubble-volume measurements. The density of the gas, required for calculating the volumetric flow rate, was evaluated for all measurements at the bath temperature. Further liquid-specific details are discussed in the Supplementary Material\ref{sec:suppmat}.\\

\begin{figure}
\centering
\includegraphics[scale=0.2]{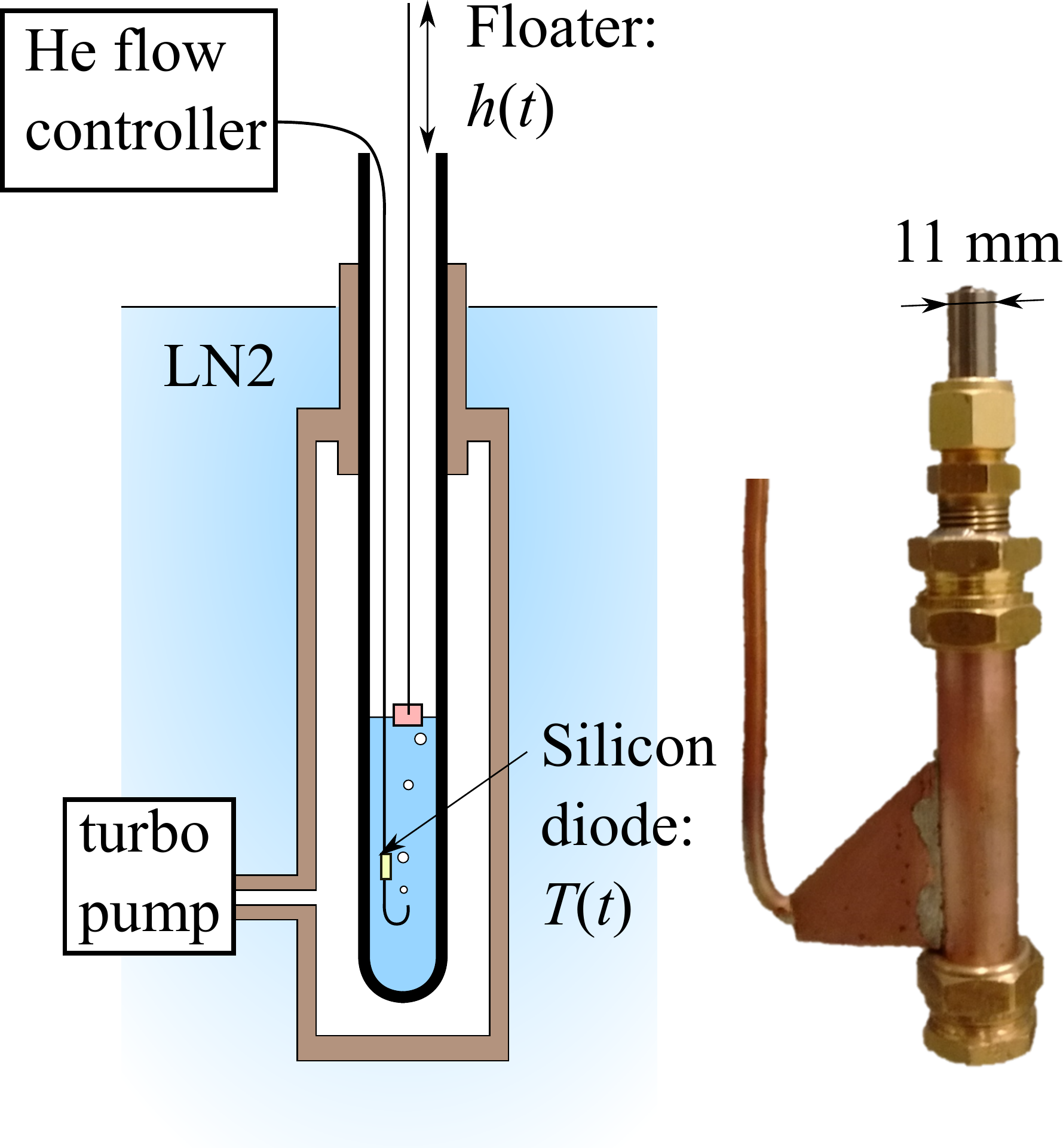}
\caption{Experimental setup for the cooling measurements. The schematic on the left is a simplified representation, whereas the picture on the right shows the realised setup without the temperature sensor and tubing for the helium gas. The level of liquid is measured by a floater on which a long glass capillary is placed. }
\label{fgr:setup}
\end{figure}

In addition to the optical measurements of the bubble growth we employed this method to study the cool down of a well defined system, see \autoref{fgr:setup}. A small amount of liquid nitrogen was filled into a steel tube, which was closed at the bottom. This geometry allowed accurate calculation of the total heat capacity for further calculations. Bubbles were generated inside this liquid bath, where the temperature and liquid level were monitored during the experiment. The liquid level was measured by means of a floater, which was equipped with a glass capillary to allow tracking its position outside the tube. To prevent heat losses, we created a vacuum around the bottom and side walls, using a second larger tube. The space created is maintained at a pressure of $\SI{e-5}{\pascal}$. Radiation was prevented by submerging the setup in a large liquid nitrogen bath, thus maintaining the outer tube and the open top of the steel tube at $T=\SI{77}{\kelvin}$

\begin{figure}
\centering
\includegraphics[scale=0.35]{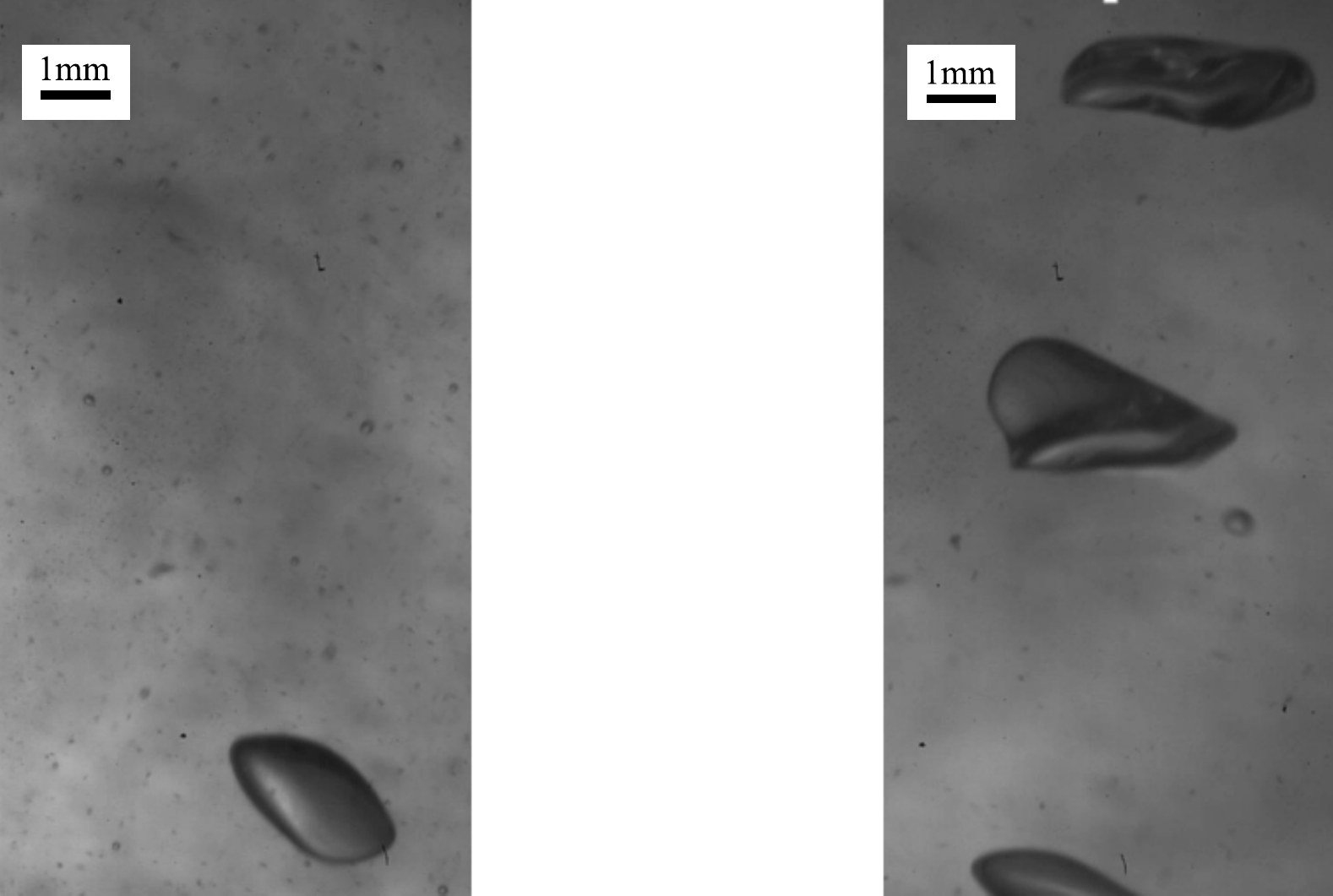}\\
\SI{71.2}{\kelvin} \hspace{0.15\textwidth} \SI{76.8}{\kelvin}
\caption{Snapshots of bubbles in a liquid nitrogen bath at two different bath  temperatures. Helium gas is injected from below (not visible) where the liquid nitrogen evaporates into the bubble, increasing the bubble size. Images are taken far away from the injection point to ensure that the final bubble size is observed.}
\label{fgr:example}
\end{figure}

As mentioned before, measurements were performed on helium gas bubbles injected in a liquid nitrogen bath and on nitrogen gas bubbles cooling a water bath. As an example, recordings for the first system are shown in \autoref{fgr:example}, where the bath temperatures were at \SI{71.2}{\kelvin} and \SI{76.8}{\kelvin}.

\begin{figure}
\centering
\includegraphics[width=0.45\textwidth]{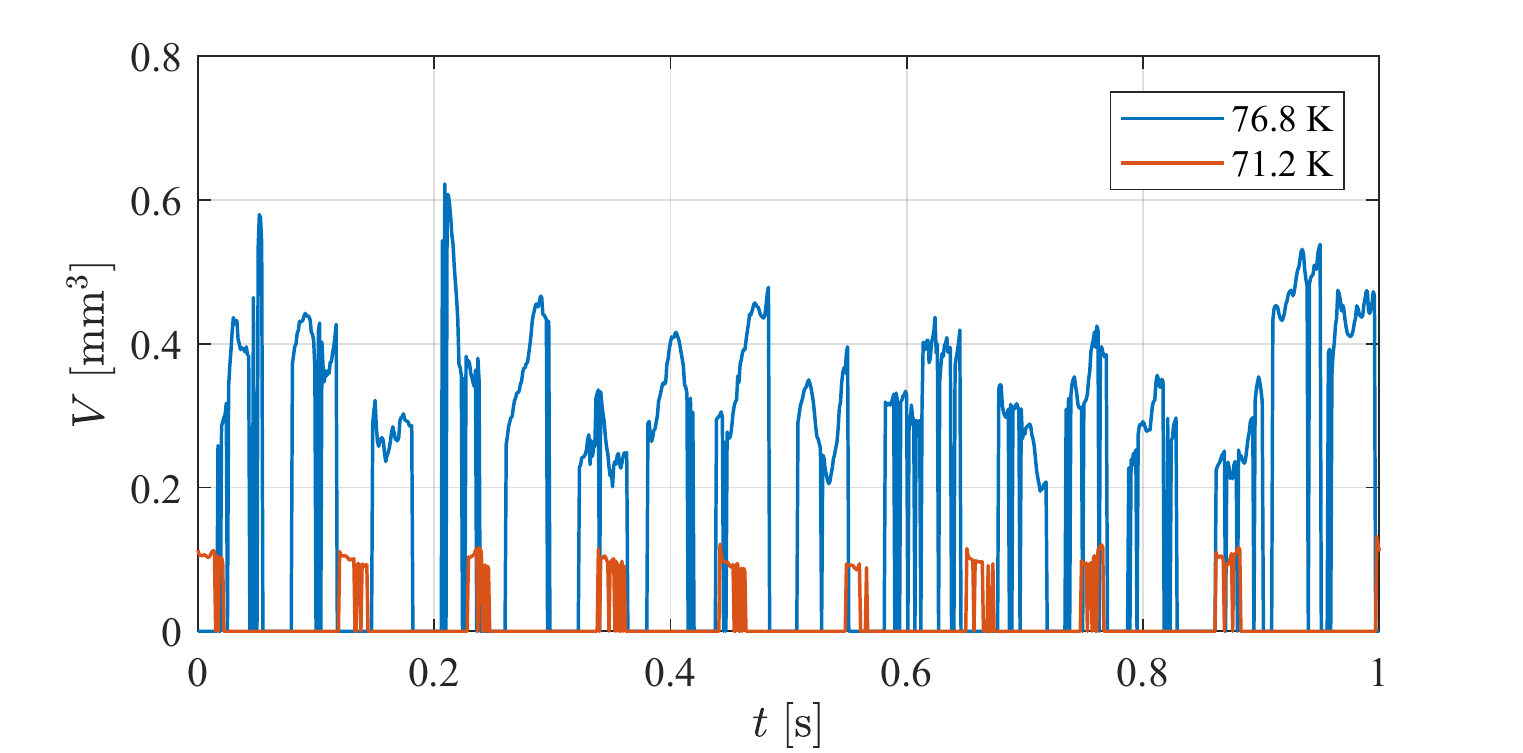}
\caption{Measurement data for helium gas bubbles injected into liquid nitrogen baths at 71.2 and \SI{76.8}{\kelvin}. Each peak represents a unique bubble. Both the bubble frequency as well as the bubble volume increases with increasing bath temperature. The increase in volume allows for larger surface deformations, reflected in fluctuations in the recorded volume.}
\label{fgr:timeseries}
\end{figure}

Using the method described in \autoref{sec:suppmat} assuming axi-symmetry, we obtain for every frame the volume of the bubbles. The (unfiltered) time-series obtained in this way are presented for the two bath-temperature cases of \autoref{fgr:example} in \autoref{fgr:timeseries}.  Comparing the two, one can readily see that the temperature of the bath is of importance: with increasing temperature, the bubble frequency is increasing as well as the size of the bubbles, which is also visible in the images of \autoref{fgr:example}.

The size and frequency difference between the bubbles for different temperatures can be understood by a pressure balance: Let $p_\mathrm{g}$ be the gas pressure (in this case, of helium) and $p_\mathrm{v}$ the vapour pressure (in this case, of nitrogen), then we obtain: 
\begin{equation}
p_\mathrm{g}+p_\mathrm{v}=p_\infty +2\sigma/R +\rho_\mathrm{l} g h,
\label{eq:p1}
\end{equation}
where $p_\infty$ is the ambient pressure of the container. The second and third terms of the right hand side denote the Laplace pressure due to the surface tension $\sigma$ and the hydrostatic pressure term (using $\rho_\mathrm{l}$ as the liquid density, $g$ the gravitational constant and $h$ the distance from the free surface of the bath). For millimetric bubbles inside a \SI{0.1}{\meter} bath, these two terms are negligible. Hence,  \autoref{eq:p1} reduces to just  $p_\mathrm{g}+p_\mathrm{v}=p_\mathrm{b} $, for which we wrote $p_\mathrm{b}=p_\infty$ as the bath pressure. This balance assumes the bubble to be at equilibrium with the liquid bath, in terms of heat as well as mass transfer. 

We measured the final bubble growth at a few centimetres from the injection point, to avoid the complex phenomena involving the dynamics. At this position the bubble has reached its final size and is at thermal equilibirium \cite{clift2005bubbles} with the well-mixed bath \cite{gvozdic2018experimental}. Note that the bath is well mixed by the rising bubbles and  recirculation inside the bubble reduces the mixing time-scales further. Additional observations revealed the growth already while the bubble was still attached  to the injection needle. \

The gas pressure can be found using the ideal gas law $p_\mathrm{g}=N k_\mathrm{B} T/V$ for the bubble before and after evaporative growth. Here $N= p_\mathrm{b} V_0/(k_\mathrm{B} T)$, the number of molecules in the initial bubble of volume $V_0$, from which we obtain for the gas pressure $p_\mathrm{g}=p_\mathrm{b} V_0/{V}$, where $V$ is the final volume of the gas-vapour bubble. Substitution into the reduced form of \autoref{eq:p1} gives the pressure balance with the saturation pressure, 
\begin{equation}
p_\mathrm{v}=p_\mathrm{b} \left(1-\frac{V_0}{V}\right)\quad \mathrm{which\; yields}\quad \frac{V}{V_0}=\left( 1- \frac{p_\mathrm{v}}{p_\mathrm{b}} \right)^{-1}.
\end{equation}
The vapour pressure is strongly temperature dependent and can be described well by the Clausius-Clapeyron equation or by  the empirical Antoine equation. When using the latter, the growth ratio results as
\begin{equation}
\frac{V}{V_0}=\left(1 - \frac{1}{p_\mathrm{b}} 10^{\left(A-\frac{B}{T_\mathrm{b}+C}\right)}\right)^{-1},
\label{eq:RrAnt}
\end{equation}
where $A,B,C$ are fluid specific constants describing the saturation pressure and $T_\mathrm{b}$ is the bath temperature. The vapour pressure according to the Clausius-Clapeyron equation equals: $p_\mathrm{v}=\exp{\left(-\frac{L}{R_\mathrm{gas}} \left( \frac{1}{T_\mathrm{b}}- \frac{1}{T_\mathrm{r}}\right)\right)}$, where the subscript $r$  denotes a reference state, $L$ the molar latent heat of evaporation and $R_\mathrm{gas}$ the universal gas constant. A further simplification can be obtained by using the saturation conditions of the substance at $p_\mathrm{r}=p_\mathrm{b}$ as the reference state, i.e. $T_\mathrm{r}=T_\mathrm{sat}(p_\mathrm{b})$, from now on abbreviated as just $T_\mathrm{s}$. As a result, the growth ratio can be expressed as
\begin{equation}
\frac{V}{V_0}=\left(1 -  \exp{}^{-\frac{L}{R_\mathrm{gas}} \left(\frac{1}{T_\mathrm{b}} - \frac{1}{T_\mathrm{s}} \right)}\right)^{-1}.
\label{eq:Rr}
\end{equation}

The degree of subcooling $-\Delta T$ equals $-(T_\mathrm{s}-T_\mathrm{b})$. We can write $\frac{1}{T_\mathrm{b}}=\frac{1}{T_\mathrm{s}}+\frac{\Delta T}{T_\mathrm{s}^2}+\mathcal{O}(\Delta T ^2)$, where the higher order terms can be neglected for relatively small $\Delta T$, resulting in \begin{equation}
 \frac{V}{V_0}=\left(1 -  \exp{}^{-\frac{L}{R_\mathrm{gas}} \frac{\Delta T}{T_\mathrm{s}^2}}\right)^{-1}.
\end{equation}

\begin{figure}
\centering
\includegraphics[width=0.45\textwidth]{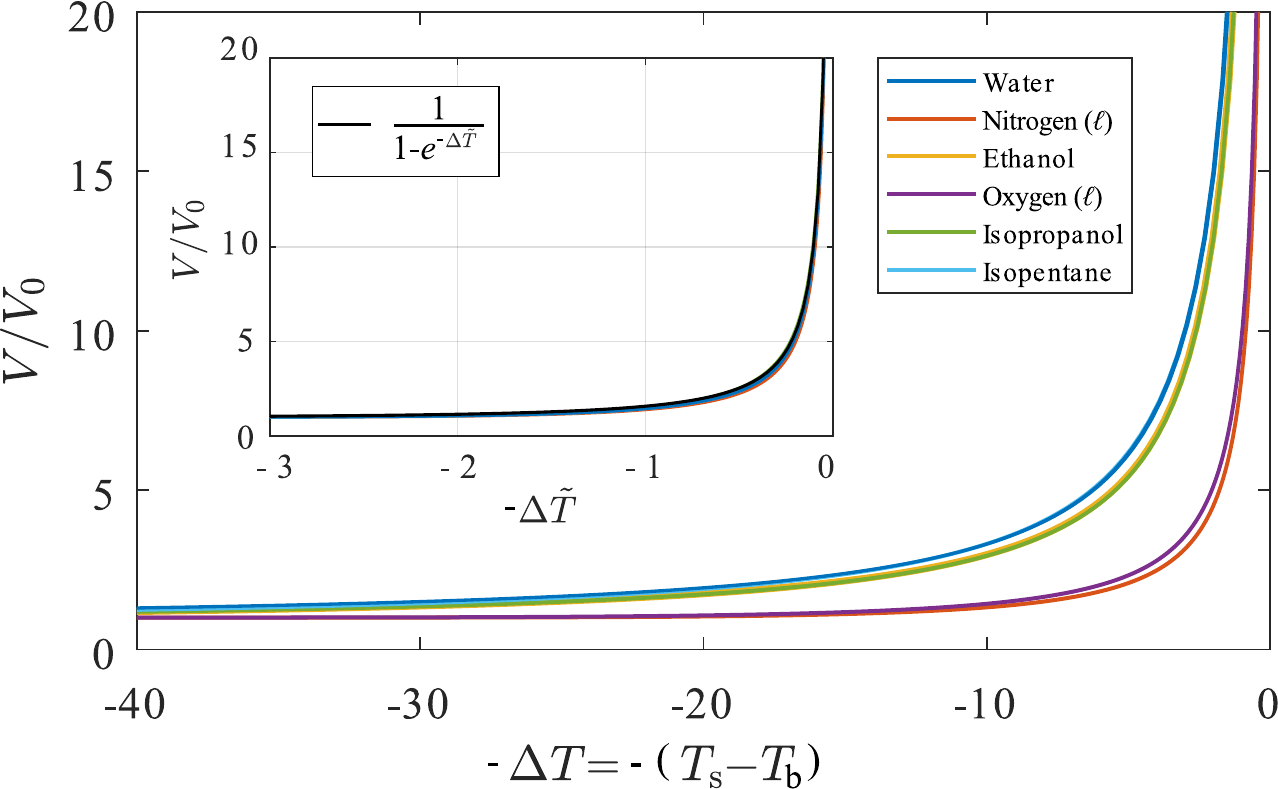}
\caption{Bubble growth ratio $\frac{V}{V_0}$ for different liquids, based on the Antoine equation with constants A,B and C taken from \cite{NIST}. Using the rescaled temperature $\Delta \tilde{T}=\Delta T\frac{R_\mathrm{gas}}{L} {T_\mathrm{s}^2}$ one can collapse all of the liquids onto a single curve, as shown in the inset. }
\label{fig:6com}
\end{figure}

When rescaling all temperatures by $\Delta \tilde{T}=\Delta T\frac{R_\mathrm{gas}}{L} {T_\mathrm{s}^2}$ we can collapse all data onto a single curve, as shown by the inset of \autoref{fig:6com}.  This provides a powerful tool to compare the bubble expansion ratios for a given temperature drop between various substances. E.g. for a bubble to expand to triple the original volume, we find $-\Delta \tilde{T}=\ln(1-1/3)\approx -0.41$. Table \ref{tbl:DT} shows the corresponding temperature drop for the six substances used in \autoref{fig:6com}, displaying good agreement.

\begin{table}
\small
\begin{tabular}{c|cccccc}
&&&&&Iso-&Iso-\\
&Water & Nitrogen&Ethanol&Oxygen&propanol&pentane\\
Eq. \ref{eq:RrAnt}&11.2&3.6&10.1&3.7&9.7&11.2\\
$\Delta \tilde{T} \frac{L}{R_\mathrm{gas} T_\mathrm{s}^2}$ &11.7&3.7&10.1&4.1&9.8&12.2
\end{tabular}
\caption{Absolute cooling temperature using the Antoine equation and the rescaled formula for a bubble expanding to three times its original volume. Antoine coefficients taken from \cite{NIST}}
\label{tbl:DT}
\end{table}

\begin{figure}
\centering
\includegraphics[width=0.45\textwidth]{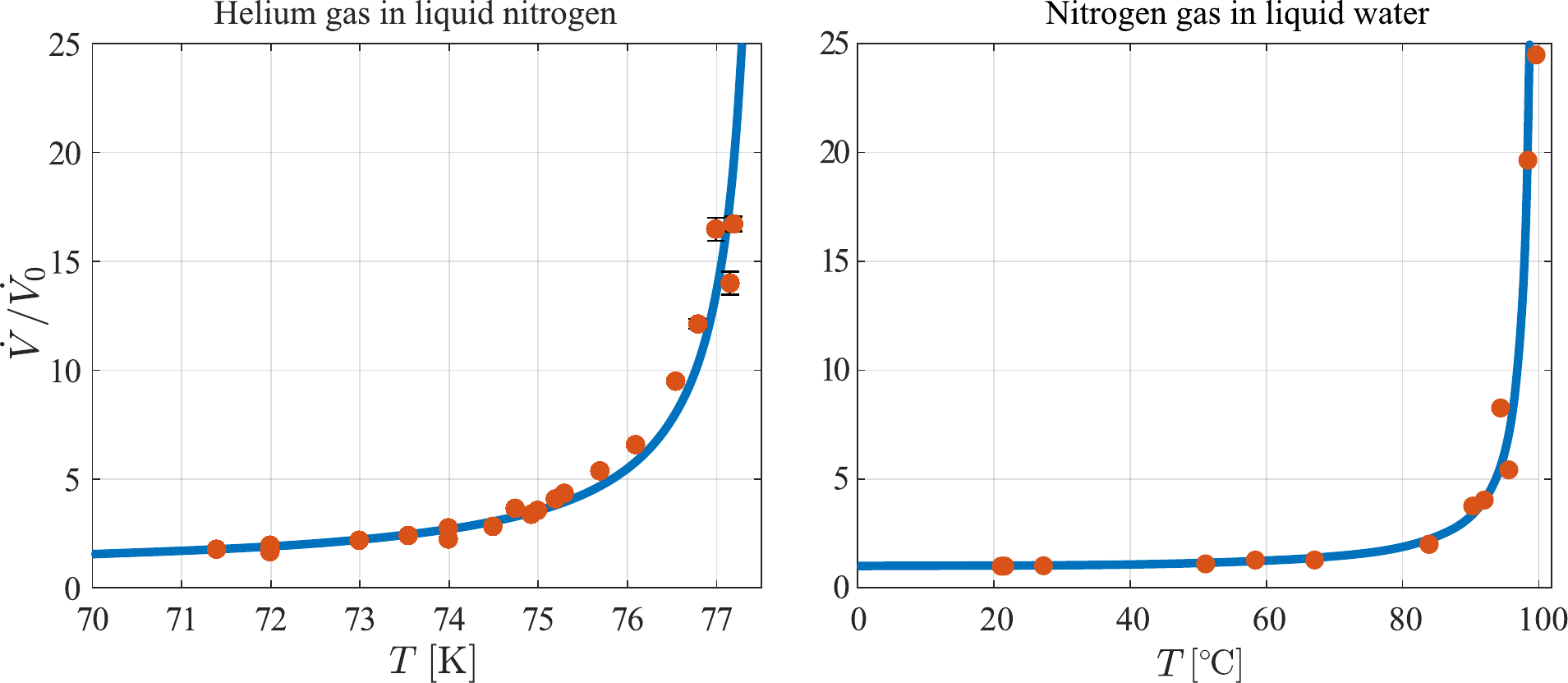}
\caption{Measurements of the ratio between the injected gas volume $\dot{V}_0$ and the (optically) observed bubble volume $\dot{V}$ passing per unit time. Two systems are studied here: helium gas injection in liquid nitrogen (left) and nitrogen gas into a water bath. Both measurements show good agreement with \autoref{eq:RrAnt} (solid line) for all temperatures. The error bars smaller than the corresponding data point are omitted}
\label{fgr:res}
\end{figure}

An indirect measure of \autoref{eq:Rr} is the observed amount of volume rising towards the free surface relative to the amount of gas being injected per unit of time, i.e. $\dot{V}_f/\dot{V}_0$, in which $_f$ and $_0$ denote the final volume and injected volume, respectively.  $V_0$ is obtained from the mass flow rate as set by the flow controller using the density of the gas at the bath temperature, whereas $\dot{V}_f$ is obtained by dividing the summed volumes of all individual bubbles passing the camera by the total time of the measurement. This way we avoid difficulties in the determination of the initial bubble size, which differs from bubble to bubble. 
We performed experimental validation of this relation by measuring the ratio in volume flux, using the setup presented \autoref{fig:SetupBare}. The results are shown in \autoref{fgr:res} for gaseous nitrogen injection into a liquid water bath and gaseous helium injection into a liquid nitrogen bath. As can be seen, the results compare well with the prediction by \autoref{eq:Rr} for both systems. 

\begin{figure}
\includegraphics[width=0.45\textwidth]{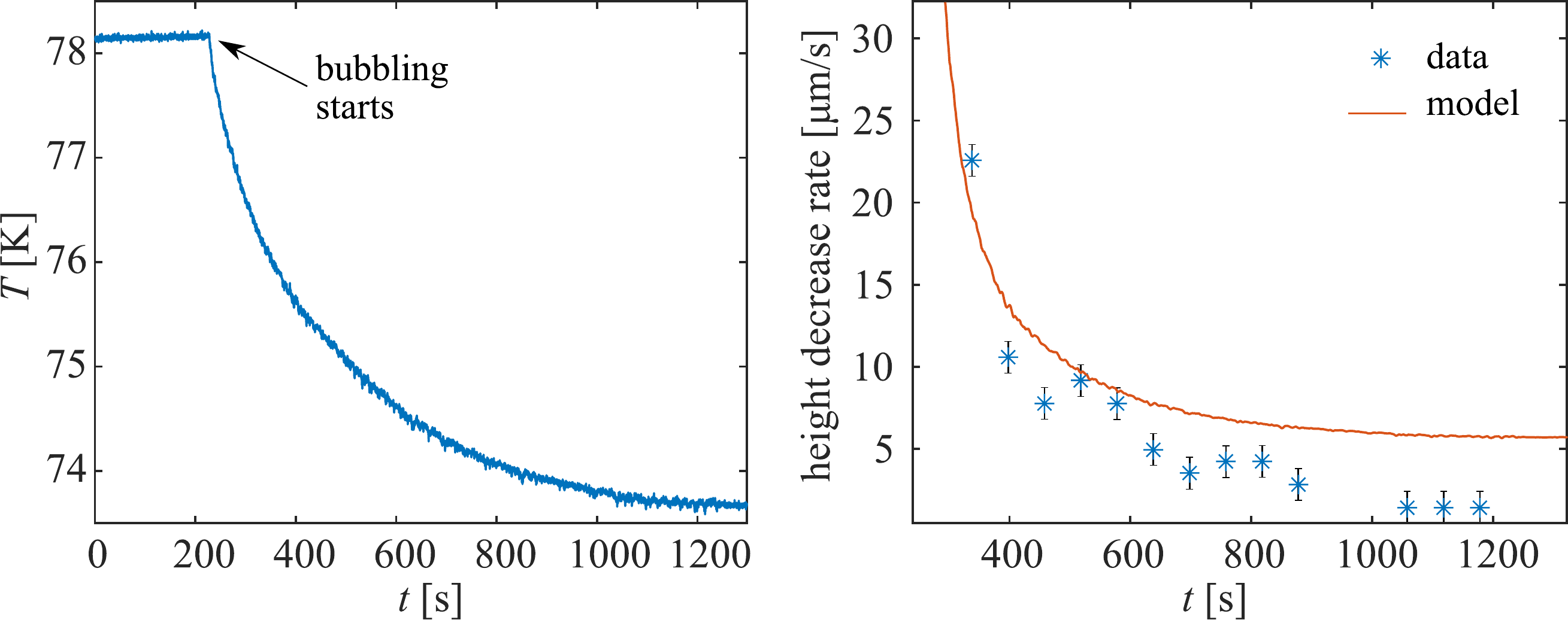}
\caption{Bubble experiment of helium gas in a  bath of \SI{4.8}{\milli \liter} liquid nitrogen. The left panel shows the temperature decrease as a result of the forced evaporation, while the right panel shows the decrease in liquid height, measured by the floater.}
\label{fig:systemTt}
\end{figure}

In a further study, the setup described in \autoref{fig:systemTt} is used to study the time evolution of a liquid volume being cooled down by forced evaporation. We prepared \SI{4.8}{\milli \liter} of liquid nitrogen and used helium gas injection to monitor the decrease in temperature $T$ and the liquid level $h$. The results are presented in \autoref{fig:systemTt}, where at $t=\SI{225}{\second}$ helium gas is injected at a rate of \SI{2e-1}{\milli \gram \per \second}. An immediate decrease in temperature can be observed, as well as a decrease in nitrogen level as a result of the forced evaporation. The decrease in temperature of the liquid can be estimated by considering the power taken from the liquid (and the container). The power can be expressed as $L \rho_\mathrm{b} (\mathrm{d}V_\mathrm{b}/\mathrm{d}t)$, which equals $L \rho_\mathrm{v} (\mathrm{d}V/\mathrm{d}t)$. The bubble expansion rate can be expressed as $ (\mathrm{d}V_0/\mathrm{d}t)(V-V_0)/V_0$, where $\mathrm{d}V_0/\mathrm{d}t$ is the volumetric flow of the helium gas $\dot{V}_0$. As a result, the rate at which the bath temperature decreases is estimated as $\mathrm{d}T/\mathrm{d}t \approx \dot{V_0} L(\frac{V}{V_0}-1)\rho_\mathrm{v}/C
$, where $C$ is the total heat capacity of the liquid nitrogen and the steel tube surrounding it. With $\dot{V}_0=\SI{3.2e-7}{\meter \cubed \per \second}$, $L=\SI{2e5}{\joule \per \kelvin}$, $\rho_\mathrm{v}=\SI{4.6}{\kilo \gram \per \meter \cubed}$ and $C=\SI{7.8}{\joule \per \kelvin}$, we find ${\Delta T}/{\Delta t}\approx \SI{e-1}{\kelvin \per \second}$ which is the right order of magnitude for the \textit{initial} cool-down rate, as can be seen from \autoref{fig:systemTt}.
 Using the established theory we can use the temperature data to calculate the expected decrease in nitrogen level and compare this with the direct measurements of the floater, see \autoref{fig:systemTt}. It can be seen that the model predicts the evaporation quite well, whereas for low temperatures an over prediction is found.  The system seems to stabilize at a temperature of about \SI{73.5}{\kelvin} at which also the liquid level hardly decreases. We attribute this to recondensation of nitrogen onto the tube wall above the liquid bath. Since helium gas has a much lower density than nitrogen, the vapour above the liquid surface will have a higher nitrogen partial pressure than the vapour in the bubble. Therefore, the nitrogen vapour will recondense at the tube wall near the surface that is at a relatively low "sub-cooling" temperature. The condensed nitrogen will flow back into the liquid bath  via the wall into the liquid  and the heat of condensation  will compensate the heat of evaporation taken from the liquid through the injection of the helium gas, negating the vapour cooling method. 

In terms of cooling power per injected gas volume we find that the bubble-cooling method excels close to the saturation temperature of the liquid, visible by the rapid decrease in bath level when the bubbling just started. This is in agreement with the visual measurements of \autoref{fgr:res}, as can be seen by the asymptotic behaviour of both the measurements and the theory. As a result of our study, the efficiency of a cooler or the mass loss in a bubble column is therefore the largest close to the saturation temperature of the bath. 
We studied the cooling behaviour of a liquid by gas injection. As a result of the \textit{initial} partial pressure being zero, the liquid evaporates into the gas bubble until equilibrium is reached.  The evaporation process takes energy from the bath, resulting in a decrease of bath temperature, as observed in our experiments. The optical measurements of the bubble growth in two different liquid-bubble pairs showed excellent agreement with our theoretical description of the phenomena. Our proposed rescaling of the temperature using liquid specific parameters allows for easy comparison and understanding of this phenomenon. It allows one to indicate the heat and mass losses in the context of chemical reactors, while predicting the efficiency for future applications using this technique as a cooling method.

\section{Supp Mat}
\label{sec:suppmat}
\paragraph{Additional experimental methods}

For the study of water, the container was made of glass and placed inside a temperature controlled copper block. Using a flow controller, the nitrogen flow was set to be $\dot{m}_0=\SI{5.2 e-2}{\milli \gram \per \second}$.

The bubble growth in liquid nitrogen was studied by helium gas injection at a flow rate of $\dot{m}_0=\SI{2.0e-2}{\milli \gram \per \second}$. It is of great importance to limit the heat losses for the setup to study liquid nitrogen. Therefore, the container was placed in a sacrificial bath, which was filled with liquid nitrogen as well. The gas flow was pre-cooled by leading the tubing through this bath. The whole setup was put inside a cryostat, which only had access at the top to prevent heat losses. As a consequence, optical measurements were performed from the top using mirrors for both the camera and the light source. An evaporator was put inside the setup to generate cold nitrogen gas to prevent icing of the setup, which otherwise would hinder optical access. Since the primary flow lacked the cooling capacity to cool down the setup, a large, secondary helium gas flow was used to cool down the bath by the same mechanism of bubble cooling.
\paragraph{Image processing}
The images from the camera were analysed using automated image processing. Firstly, edge detection was used by means of a 'Canny' filtering technique \cite{Canny1986}. The resulting binary image contains various regions. When choosing the proper filtering parameters, the largest of these regions are the bubbles for which the position and volume has to be calculated.  The volume was obtained by first slicing the area into horizontal lines from which the center position $r_\mathrm{c}(z)$ and radius $R(z)$ can be determined. Assuming axi-symmetry, this allows us to express the bubble volume 
$
V=2 \pi\int^h_0 (R(z)-r_\mathrm{c}(z))^2 \mathrm{d}z,
$
which was implemented numerically using the trapezoid integration method.

\paragraph{Post processing}
Since the bubble has a size of roughly the capillary length $(\sigma/\rho_\mathrm{l} g)^{1/2}$ surface deformations are expected. This effect is amplified by the increasing velocity and size of the bubble during the rise towards the free surface of the bath. Therefore, the bubble is no longer axi-symmetric and the apparent volume fluctuates between frames. Temporal averaging was therefore performed for every bubble to obtain a more accurate bubble volume.

\paragraph{Error analysis}
Multiple systematic errors can be identified, which are found to be negligible: Calibration error of the camera, uncertainties in the temperature probe and gas flow controller. The largest error in our measurements, however, emerges from the aforementioned bubble shape fluctuations. Therefore, we used the variance in measured bubble size as a statistical error: $\sigma^2=\Sigma_j \sigma^2_j$, where $\sigma^2_j$ indicates the variance for every bubble $_j$. The second source of error lies within the accuracy of the temperature measurements, which is $\SI{0.2}{\kelvin}$.

\bibliographystyle{ieeetr}

\end{document}